\documentclass[aps,pra,showpacs,twocolumn,amsmath,amssymb,superscriptaddress]{revtex4-1}
\usepackage{graphicx}
\begin{document}
%\begin{CJK*}{GBK}{song}
\title{Defect Solitons in Parity-Time Symmetric Optical Lattices with Nonlocal Nonlinearity}

%% For REVTeX it is possible to automate superscript and e-mail callouts with the superscriptaddress option; see REVTeX4 documentation.

\author{Sumei Hu}
\affiliation{Laboratory of Nanophotonic Functional Materials and Devices, South China Normal University,  Guangzhou  510631, P. R. China}
\affiliation{Department of physics, Guangdong University of Petrochemical Technology, Maoming 525000, P. R. China}
\author{Xuekai Ma}
\affiliation{Laboratory of Nanophotonic Functional Materials and Devices, South China Normal University,  Guangzhou  510631, P. R. China}
\author{Daquan Lu}
\affiliation{Laboratory of Nanophotonic Functional Materials and Devices, South China Normal University,  Guangzhou  510631, P. R. China}
\author{ Yizhou Zheng}
\affiliation{Laboratory of Nanophotonic Functional Materials and Devices, South China Normal University,  Guangzhou  510631, P. R. China}
\author{ Wei Hu}\email[Corresponding author's email address:
]{huwei@scnu.edu.cn}
\affiliation{Laboratory of Nanophotonic Functional Materials and Devices, South China Normal University,  Guangzhou  510631, P. R. China}

\begin{abstract}
The existence and stability of defect solitons in parity-time (PT) symmetric  optical lattices with nonlocal nonlinearity are reported.
It is found that nonlocality can expand the stability region of defect solitons. For positive or zero defects, fundamental and  dipole solitons can exist stably in the semi-infinite gap and the first gap, respectively.
For  negative defects, fundamental solitons can be stable in both the
semi-infinite gap and the first gap, whereas dipole
solitons are unstable in the first gap. There exist a maximum degree of nonlocal nonlinearity, above which the fundamental solitons in the semi-infinite gap and the dipole solitons in the first gap do not exist for negative defects.
The influence of the imaginary part of the PT-symmetric potentials on soliton stability is given.
When the modulation depth of the PT-symmetric lattices is small, defect solitons can be stable for positive and zero defects, even if the PT-symmetric potential is above the phase transition point.
\end{abstract}

%\ocis{(190.0190) nonlinear optics; (190.6135) Spatial solitons. }% REPLACE WITH CORRECT OCIS CODES FOR YOUR ARTICLE
                          % NOTE: \ocis{} IS ALIASED TO \pacs{} BUT MUST
                          % FORMAT THE TERMS CORRECTLY FOR EACH JOURNAL

\maketitle %% null function with osajnl.sty
%\end{CJK*}

%%%%%%%%%%%%%%%%%%%%%%%%%%  body  %%%%%%%%%%%%%%%%%%%%%%%%%%
\section{Introduction}
The physical systems with parity-time (PT) symmetry have attracted much
more attention in recent years
\cite{Bender-1999,Bender-2003,El-Ganainy2007-ol,PT2008-PRL030402,PT2008-PRL103904,PT2010-PRL,PT2009-PRL,New2011-pra,nature-phys2010,PT2009-prl093902}.
Bender {\it et al. } found that PT-symmetric Hamiltonians can have entirely real
spectra although these Hamiltonians are non-Hermitian \cite{Bender-1999,Bender-2003}.
In optics, PT-symmetric structures can be constructed by inclusion of gain or loss regions
into  waveguides, which make the complex refractive-index
distribution obeying the condition $n(x)=n^*(-x)$
\cite{El-Ganainy2007-ol,PT2008-PRL030402,PT2008-PRL103904}.
%Dissipative periodic waves, solitons and breathers in nonlinear Schr\"odinger equation
%with complex potentials has been studied by the authors\cite{Dissipative-pre056606}.
Many unusual features stemming from the PT-symmetry have been found, such as power oscillation
\cite{PT2008-PRL103904,nature-phys2010,PT2010-Zheng}, absorption
enhanced transmission \cite{PT2009-prl093902},  nonlinear switching
structure \cite{New-PRA2010,PT2010-Sukhorukov}, and unidirectional invisibility
\cite{PT2011-Zin}. In experiments, the PT-symmetry breaking in
complex optical potentials was observed  firstly  by Guo {\it et al}
in a double-well structure fabricated through a multilayer
$Al_{x}Ga_{1-x}As $ heterostructure with varying
concentrations\cite{PT2009-prl093902}. R\"{u}ter {\it et al}
observed the non-reciprocal wave propagation in an active
PT-symmetric coupled waveguide system based on Fe-doped $LiNbO_{3}$,
in which optical gain is provided through two-wave mixing using the
photorefractive nonlinearity \cite{nature-phys2010}.
Those theoretical and experimental results led to the proposal of a
new class of PT-symmetric synthetic materials with intriguing and
unexpected properties that rely on the non-reciprocal light propagation
\cite{nature-phys2010,PT2011-Zin}.

In optics, nonlinearities in the PT-symmetric systems have been
considered by many authors
\cite{PT2010-Sukhorukov,New2011-pra,PT2011-Zin,PT2011-Abdullaev,PT2011-Zezyulin},
especially in the PT-symmetric optical lattices
\cite{PT2008-PRL030402,PT2008-PRL103904,PT2011-Zin}, and some new
kinds of soliton were found and investigated
\cite{PT2008-PRL030402,DS2011-ol,OL2010-Zhou,DPT2011-OE, Lu2011-OE}.
Defect solitons can be formed when a local
defect is introduced into the optical lattices,
which have been widely studied in the regimes of photonic crystals,
waveguide arrays, and optically induced photonic lattices
\cite{DS2008-PRA,DS2009-pra,DS2006-prl}.
Due to its unique properties based on the defect guiding phenomena which can be controlled through variation of the defect parameters \cite{DS2005-prl}, defect solitons have the potential
applications for the all-optical switches and routing of optical
signals \cite{DS2004-PHT}. Zhou {\it et al} have studied the defect
modes in the PT-symmetric optical lattices \cite{OL2010-Zhou}.
Defect solitons in PT-symmetric optical lattices and superlattices
with local nonlinearity have been studied, and stable solitons are
found mainly in the semi-infinite gap \cite{DPT2011-OE,Lu2011-OE}.

It is noteworthy that the nonlinearity in the photorefractive media, in which R\"{u}ter {\it et al.} have observed the non-reciprocal wave propagation \cite{nature-phys2010}, is nonlocal due to the diffusion mechanism of charge carriers \cite{Segev1992-prl,Segev1994-prl,Segev1998-prl}. The nonlocality of
nonlinear response exists in many real physical systems, such as
photorefractive crystals \cite{Segev1992-prl,Segev1994-prl,Segev1998-prl} , nematic liquid
crystals \cite{Peccianti2002-OL,conti2003-prl,Conti2004-prl}, lead glasses \cite{Rotschild2005-prl,Alfassi2007-prl}, etc.
The nonlocality can drastically modify the properties of
solitons and improve the stability of
solitons \cite{Science1997-Snyder,prl2005-Xu,Buccoliero2007-PRL}.  Therefore it is worthy to study the properties
of solitons in PT-symmetric optical lattices with nonlocal nonlinearity.

In this paper, we study defect solitons in PT-symmetric optical
lattices with nonlocal nonlinearity. The properties of nonlocal defect
solitons are quite different from those in local media, and the
nonlocality can expand the stability regions of soliton, especially in the
first gap. It is found that the stability of nonlocal defect
solitons depends on the defect, the degree of nonlocality, and the PT-symmetric
potentials. The influence of the imaginary part of the PT-symmetric potentials is studied, and the example of stable soliton in the PT-symmetric potentials above the phase transition point is given.

\section{Theoretical Model}
We consider the propagation of light beam in the PT-symmetric defective lattices with  Kerr-type  nonlocal nonlinearity. The evolution of complex amplitude $U$ of the light fields can be described by following dimensionless nonlinear Schr\"{o}dinger equation (NLSE),
\begin{equation}\label{solution}
i\frac{\partial U}{\partial z}+\frac{1}{2}\frac{\partial^2 U}{\partial x^2}
 +p[V(x)+i W(x)]U+nU= 0,
\end{equation}
where $x$ and $z$ are the transverse and longitudinal coordinates, respectively, and $p$ is the depth of the PT-symmetric potentials. $V(x)$ and $W(x)$ are the real and imaginary parts of the PT-symmetric potentials, respectively, which are assumed in this paper as
\begin{eqnarray}\label{realpotentials}
 V(x)&=&\cos^2(x)[1+\epsilon\exp(-x^8/128)],\\
  W(x)&=&W_0\sin(2x).
\end{eqnarray}
Here $\epsilon$ represents the strength of the defect, which is expressed as a super-Gaussian profile \cite{DPT2011-OE}. The parameter $W_0$ represents the strength of the imaginary part of the PT-symmetric potentials compared with the real part.

The nonlinear refractive-index change $n$ satisfies
\begin{equation}\label{nonlinear}
n-d\frac{\partial^2 n}{\partial x^2}=|U|^2,
\end{equation}
where $d$ stands for the nonlocality degree of the nonlinear
response.  This type of nonlinear response with a finite region of nonlocality exists in many real physical systems, for instance, all diffusion-type nonlinearity
\cite{Segev1998-prl,Conti2007-prl,Conti2009-PRL}, orientational-type nonlinearity
\cite{Peccianti2002-OL,Hu2006-APL}, and the general quadratic nonlinearity
describing parametric interaction \cite{Nikolov2003-prl,Larsen2006-PRE}.
Although the diffusion nonlinearity in photorefractive media is anisotropic,
Eq.~(\ref{nonlinear}) can still be used to describe its nonlocal feature theoretically
when we ignore the anisotropic effect \cite{newnonlocal}.
% for example, the Fe-doped $LiNbO_{3}$ crystals used in the experimental observation of parity-time symmetry in optics  \cite{nature-phys2010},  can also be described by this equation.
Moreover when $d\rightarrow 0$, the above equation reduces to
$n=|U|^2$, and Eq. (\ref{solution})  reduces to that for local case.
When $d \neq 0$, Eqs. (\ref{solution}) and (\ref{nonlinear}) denote
a nonlocal NLSE with an exponential-decay type nonlocal response.

\begin{figure}[htbp]
   \centering
   \includegraphics[width=7.0cm] {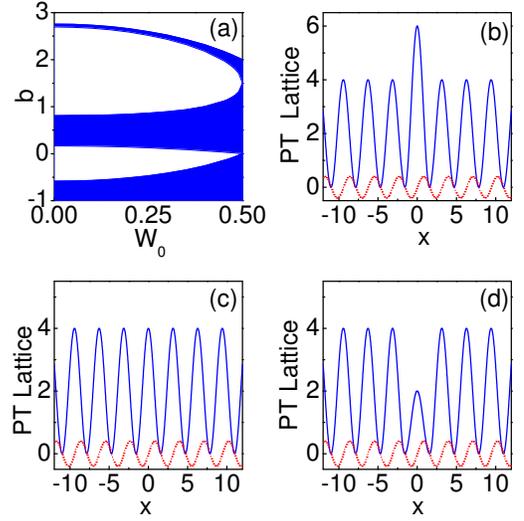}
   \caption{(color online) (a) Band structure for the PT-symmetric lattice with $p=4$; The blue regions are bands while the blank regions are gaps.
   (b)-(d) Lattice profiles for lattices with $p=4$, $W_0=0.1$, and (b) $\epsilon=0.5$, (c) $\epsilon=0$, and (d) $\epsilon=-0.5$, respectively. Solid blue and dotted red lines represent the real and imaginary parts, respectively.}  \label {Fig1}
 \end{figure}

The PT-symmetric lattices governed by Eq.~(\ref{realpotentials}) have  Bloch band
structures when $\epsilon=0$. The band diagram can be
entirely real when the system is operated below the phase transition
point ($W_0^c=0.5$) \cite{PT2008-PRL030402,PT2008-PRL103904}. The Bloch band structure obtained by the plane wave expansion method for $p=4$ is shown in
Fig.\ref{Fig1}(a). From Fig. \ref{Fig1}(a), we can see that the
region of the first gap decreases with increasing  $W_0$. The first gap disappears when $W_0=0.5$, which is the  phase transition point. When the system is above the phase transition point,  i.e. $W_0>0.5$, the band structure becomes complex. Figures
\ref{Fig1}(b)-\ref{Fig1}(d) show the profiles of the PT-symmetric potentials for
$p=4$ and $W_0=0.1$ with positive defects ($\epsilon=0.5$), zero defects
($\epsilon=0$), and  negative defects ($\epsilon=-0.5$),
respectively.

We search for stationary solutions to Eqs. (\ref{solution}) and (\ref{nonlinear}) in the form $U=f(x)\exp(ibz)$, where  $f(x)$ is a complex function satisfies equations,
\begin{eqnarray}
 bf=\frac{1}{2}\frac{\partial^2 f}{\partial x^2} +p[V(x)+i W(x)]f+nf, \label{field}
\\
 n - d \frac{\partial^2 n}{\partial x^2}=|f|^2. \label{nonfield}
\end{eqnarray}
The solutions of defect solitons are gotten numerically by the
modified squared-operator method \cite{yang-2007} from Eqs. (\ref{field}) and
(\ref{nonfield}) and  shown in the next section. To
elucidate the stability of defect solitons, we search for perturbed
solution to Eqs. (\ref{solution}) and (\ref{nonlinear}) in the form
{ $U(x,z)=[f(x)+u(x,z)+i v(x,z)] \exp(ibz)$}, where the real
[$u(x,z)$] and imaginary [$v(x,z)$] parts of the perturbation can
grow with a complex rate $\delta$ upon propagation. Linearization of
Eq. (\ref{solution}) around the stationary solution $f(x)$ yields
the  eigenvalue problem,
\begin{eqnarray} \label{solution7}
 \delta v& =& \frac{1}{2}\frac{\partial^2 u}{\partial x^2}+ (n-b) u  +p(V u-W v)\nonumber \\ &+&Re[f(x)]\int_{-\infty}^{\infty}2 G(x-\xi)u(\xi)Re[f(\xi)]{\rm d}\xi \nonumber \\
&  + & Re[f(x)] \int_{-\infty}^{\infty} 2 G(x-\xi)Im[f(\xi)] v(\xi) {\rm d}\xi,
\\
 \delta u & = & -\frac{1}{2}\frac{\partial^2 v}{\partial x^2} -(n-b) v  -p(Wu+Vv) \nonumber \\
 & - &Im[f(x)]  \int_{-\infty}^{\infty} 2G(x-\xi)Im[f(\xi)] v(\xi) {\rm d} \xi \nonumber \\
& -& Im[f(x)]\int_{-\infty}^{\infty} 2G(x-\xi)Re[f(\xi)] u(\xi) {\rm d}\xi.
\end{eqnarray}
Here $G(x)=[1/(2d^{1/2})]\exp(-|x|/d^{1/2})$ is the response function of nonlocal nonlinearity. Above  eigenvalue problem is solved numerically to find the maximum value of $Re(\delta)$. If $Re(\delta)>0$, solitons are unstable. Otherwise, they are stable.

\section{Nonlocal Defect Solitons}
In the PT-symmetric defective lattices with  nonlocal nonlinearity, we find two types
of defect solitons for positive, zero, and negative defects,
respectively. The first type is the nodeless fundamental solitons, which
can exist stably in the semi-infinite gap for all kinds of defects
or in the first gap for negative defects. The other type of
defect solitons, which exist in the first gap for  all kinds of defects, is called dipole
solitons in this paper, because most of them have two significant intensity peaks.

\begin{figure}[htbp]
    \centering
    \includegraphics[width=6.9cm]{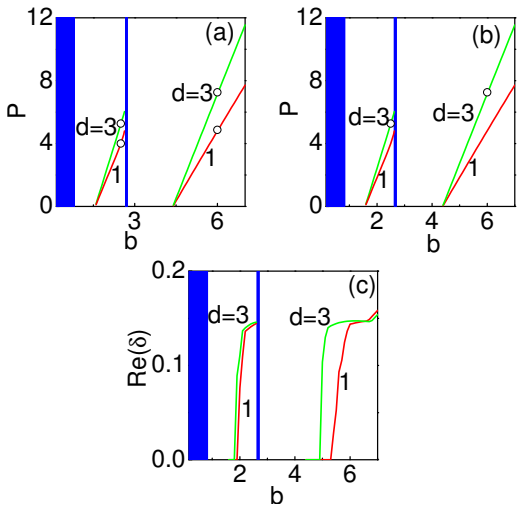}
    \caption{(color online) Soliton power versus propagation constant for positive defect with (a) $W_0=0.1$ and (b) $W_0=0.15$.
    (c) The unstable growth rate $Re(\delta)$ for soliton solutions in (b). For all cases $\epsilon=0.5$ and $p=4$.} \label{Fig2}
\end{figure}

\begin{figure}[htbp]
    \centering
    \includegraphics[width=6.9cm]{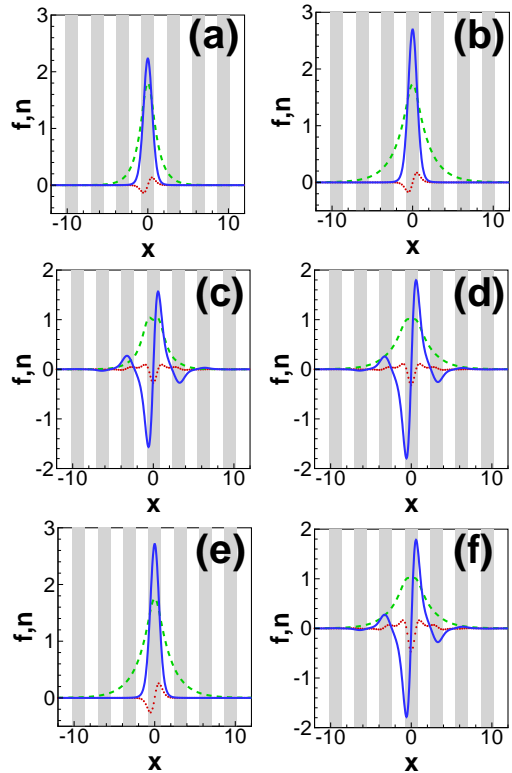}
    \caption{(color online) The complex fields (solid blue: real part; dotted red: imaginary part) and the refractive-index changes (dashed green) for soliton solutions at (a) $b=6$, $d=1$; (b) $b=6$, $d=3$; (c) $b=2.5$, $d=1$;  (d) $b=2.5$, $d=3$; (e) $b=6$, $d=1$; and (f) $b=2.5$, $d=3$; respectively. For all cases  $\epsilon=0.5$, $p=4$. For (a)-(d) $W_0=0.1$ and (e)-(f) $W_0=0.15$.} \label{Fig3}
\end{figure}

\begin{figure}[htbp]
  \centering
  \includegraphics[width=6.6cm]{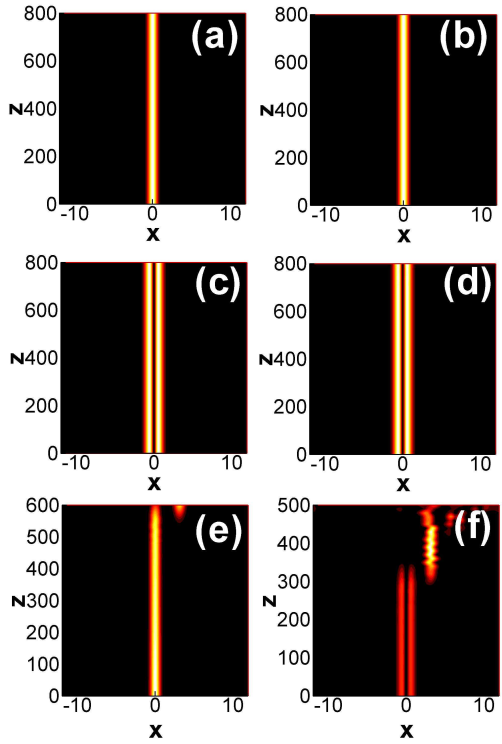}
  \caption{(color online) (a)-(f) Evolutions of defect solitons corresponding to those in Figs. \ref{Fig3}(a)- \ref{Fig3}(f), respectively.} \label{Fig4}
  \end{figure}

For positive defects, we assume $\epsilon=0.5$ and the results
are shown in Figs. \ref{Fig2} - \ref{Fig4}. Figures \ref{Fig2}(a) and \ref{Fig2}(b) show that the power of soliton [defined as {$P=\int_{-\infty }^{+\infty}|f(x)|^2
{\rm d}x$}] for both fundamental and dipole solitons increases
almost linearly as increasing of propagation constant. As the nonlocality degree $d$ increasing, the power of soliton increases too. The positive defect
solitons vanish when  the propagation constant is below the cutoff
point, whose value does not depend on the nonlocality degree. This
feature is similar to the case of traditional uniform lattices
in nonlocal media \cite{prl2005-Xu}.

When $W_0=0.1$, the linear stability analysis shows that both
fundamental and dipole solitons are stable in their whole
regimes of existence in the semi-infinite gap and the first
gap, respectively. For comparison, dipole solitons in the PT-symmetric
defective lattices with local nonlinearity are stable only in a small region near the edge
of the first gap \cite{DPT2011-OE}. It shows that the nonlocality
expands the stability region of soliton. As $W_0$ increases, the stability region of defect soliton decreases. Figure \ref{Fig2}(c) shows the maximum growth
rate $Re(\delta)$ for solitons with  $W_0=0.15$. We can
see that both fundamental and dipole solitons are unstable
for the large propagation constant, and their stability regions become very narrow.
We find that all fundamental solitons are unstable and their stability region disappears when $W_0>0.24$.  For the dipole solitons in the first gap,  the maximum value for existence of stable soliton is $W_0=0.16$.


The field profiles of the positive defect solitons are shown
in Figs. \ref{Fig3}(a)-\ref{Fig3}(f) for different $W_0$ and
nonlocal degrees $d$, which correspond to the circle symbols in
Figs. \ref{Fig2}(a) and \ref{Fig2}(b). Figures \ref{Fig3}(a) and \ref{Fig3}(b)
show fundamental solitons in the semi-infinite gap for $W_0=0.1$, while dipole
solitons in the first gap are shown in Figs. \ref{Fig3}(c) and
\ref{Fig3}(d). One can see that the poles of dipole solitons are located inside the
central channel of the lattice. As the nonlocality degree increases, the
amplitudes of solitons increase but their shapes change very little.  Figures \ref{Fig3}(e) and \ref{Fig3}(f) show the defect solitons for $W_0=0.15$.
As the $W_0$ increases, the imaginary part of the field increases.  The propagations corresponding to those solitons in Figs. \ref{Fig3}(a)-\ref{Fig3}(f) are shown in
Figs. \ref{Fig4}(a)-\ref{Fig4}(f). The evolutions are simulated based on Eqs.
(\ref{solution}) and (\ref{nonlinear}), and 1\% random-noise
perturbations are added into the initial input to verify the results
of linear stability analysis. We can see that both fundamental
solitons and dipole solitons are stable for $W_0=0.1$ but unstable
for $W_0=0.15$.

\begin{figure}[htbp]
    \centering
    \includegraphics[width=6.9cm]{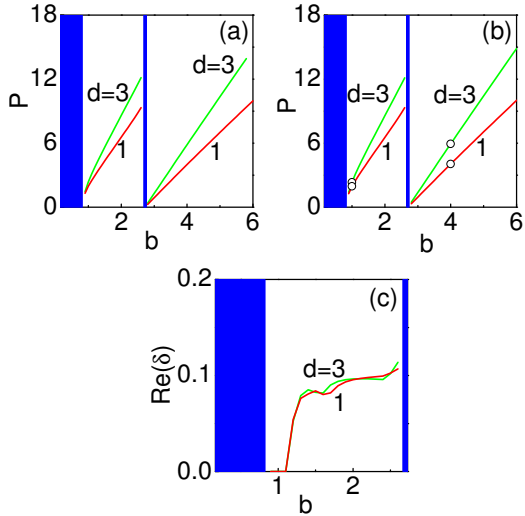}
    \caption{(color online) Soliton power versus propagation constant for zero defect with (a) $W_0=0.05$ and (b) $W_0=0.1$, respectively.
    (c) The unstable growth rate $Re(\delta)$ for soliton solutions in (b). For all cases $\epsilon=0$ and $p=4$.} \label{Fig5}
\end{figure}

\begin{figure}[htbp]
    \centering
    \includegraphics[width=6.9cm]{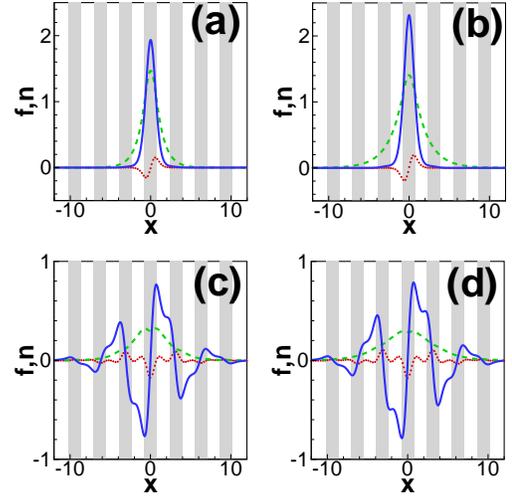}
    \caption{(color online) The complex fields (solid blue: real part; dotted red: imaginary part) and the refractive-index changes (dashed green) for solitons in the semi-infinite gap at (a) $b=4$, $d=1$; (b) $b=4$, $d=3$; or solitons in the first gap  at (c) $b=1$, $d=1$; and (d) $b=1$, $d=3$, respectively. For all cases $\epsilon=0$, $p=4$ and $W_0=0.1$.} \label{Fig6}
\end{figure}

Figures \ref{Fig5} and \ref{Fig6} show the results for zero
defects ($\epsilon=0$). The properties of defect solitons with zero defects are similar to those with positive defects. From Figs.
\ref{Fig5}(a)  and \ref{Fig5}(b), we can see that the cutoff points
approach the lower edges of gaps. For $W_0=0.05$, fundamental solitons  are stable in
the whole semi-infinite gap, whereas dipole solitons are stable in the whole
first gap. When $W_0=0.1$,  dipole solitons are
unstable for the large propagation constant, and their stability region becomes  narrow, as shown in Fig. \ref{Fig5}(c). The stability region of dipole solitons vanishes when $W_0>0.14$.  Fundamental solitons are still stable in the whole semi-infinite gap for $W_0=0.1$, and their stability region begins to reduce when $W_0>0.12$. When $W_0>0.23$,  all fundamental solitons become unstable. Figure \ref{Fig6} shows
the complex profiles of solitons for zero defects, which correspond to the circle symbols in
Fig. \ref{Fig5}(b). Although the examples of solitons in the first gap [Figs. \ref{Fig6}(c) and \ref{Fig6}(d)] have more intensity peaks, the most of solitons in the first gap still have  two significant peaks. The propagations of those examples  for $W_0=0.1$ in Fig. \ref{Fig6} are all stable as the expectation by the linear stability analysis.

\begin{figure}[htbp]
    \centering
    \includegraphics[width=7.0cm]{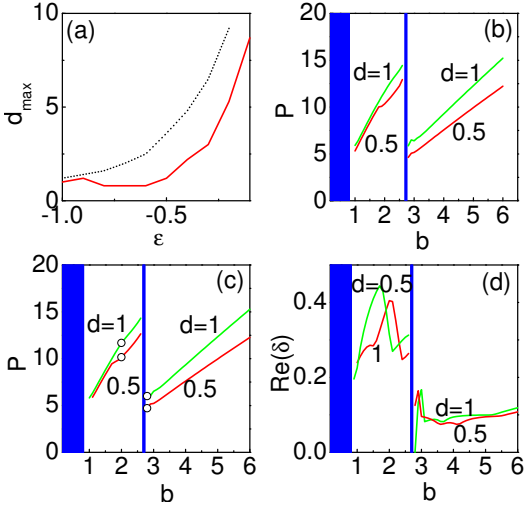}
    \caption{ (color online) (a) The maximum nonlocality degree $d_{max}$ versus the depth of negative defect for fundamental solitons in the semi-infinite gap (dashed line) and dipole solitons in the first gap (solid red line).
(b) and (c) Soliton power versus propagation constant for negative defects with (b) $W_0=0.05$ and (c) $W_0=0.1$, respectively.
    (d) The unstable growth rate $Re(\delta)$ for soliton solutions in (c). For all cases  $p=4$ and for (b)-(d) $\epsilon=-0.5$.} \label{Fig7}
\end{figure}
\begin{figure}[htbp]
    \centering
    \includegraphics[width=6.8cm]{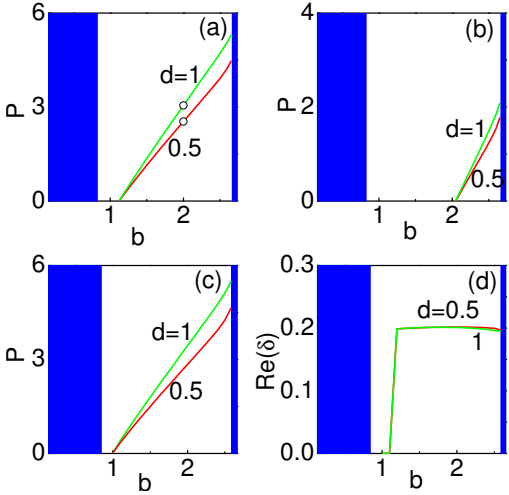}
    \caption{(color online) Soliton power versus propagation constant for fundamental solitons in the first gap with (a) $W_0=0.1$, $\epsilon=-0.5$; (b) $W_0=0.1$, $\epsilon=-0.2$; and (c) $W_0=0.2$, $\epsilon=-0.5$, respectively.
    (d) The unstable growth rate $Re(\delta)$  for soliton solutions in (c). For all case $p=4$.} \label{Fig8}
\end{figure}

\begin{figure}[htbp]
    \centering
    \includegraphics[width=7.0cm]{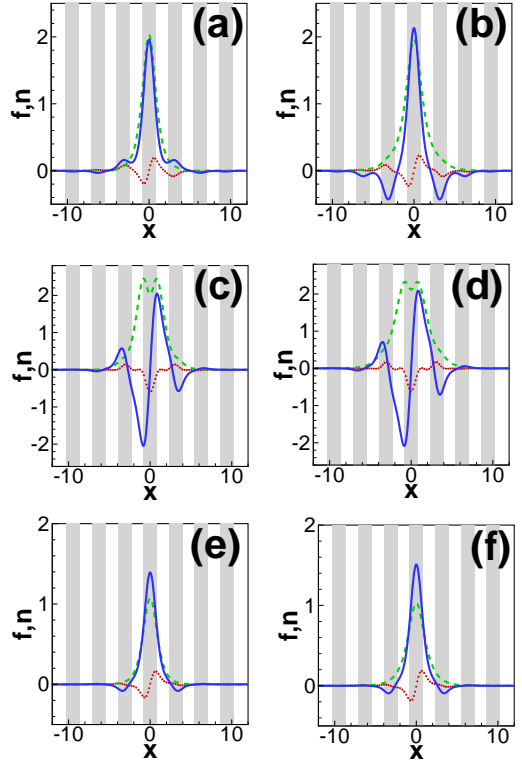}
    \caption{(color online) The complex fields (solid blue: real part; dotted red: imaginary part) and refractive-index changes (dashed green) for soliton solutions in the semi-infinite gap at (a) $b=2.8$, $d=0.5$; (b) $b=2.8$, $d=1$; or soliton solutions in the first gap at (c) $b=2$, $d=0.5$; (d) $b=2$, $d=1$;  (e) $b=2$, $d=0.5$;  and (f) $b=2$, $d=1$, respectively. For all cases $\epsilon=-0.5$, $p=4$ and $W_0=0.1$.} \label{Fig9}
\end{figure}

\begin{figure}[htbp]
  \centering
  \includegraphics[width=6.9cm]{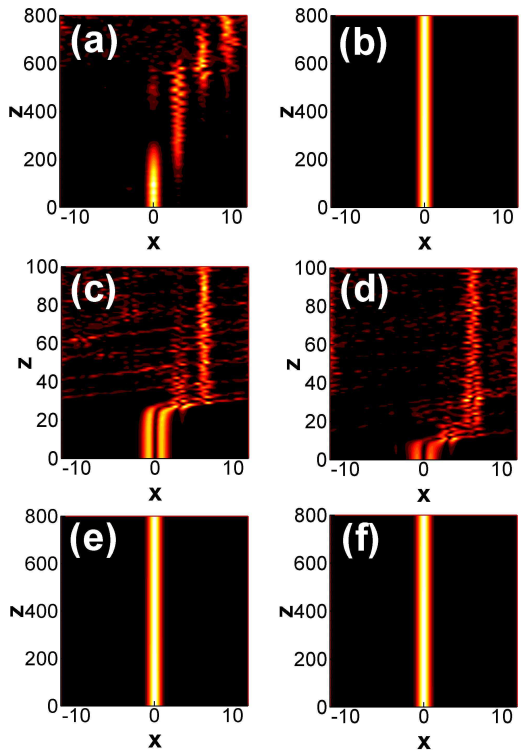}
  \caption{(color online) (a)-(f) Evolutions of defect solitons corresponding to those in Figs. \ref{Fig9}(a)- \ref{Fig9}(f) respectively.} \label{Fig10}
  \end{figure}

Next we study defect solitons in the nonlocal PT-symmetric
lattices with  negative defects, and results are
shown in Figs. \ref{Fig7}- \ref{Fig10}.  Fundamental solitons are found both in the semi-infinite gap [Fig. \ref{Fig7}] and the first gap [Fig. \ref{Fig8}], whereas dipole solitons exist in the  first gap [Fig. \ref{Fig7}].
It is interesting that there exists a maximum nonlocality degree $d_{max}$ for some negative defect solitons. When the degree of nonlocal nonlinearity is above the maximum value, the fundamental solitons in the semi-infinite gap and the dipole solitons in the first gap do no longer exist.  The maximum nonlocality degree depends on the depth of  negative defects, as shown in Fig. \ref{Fig7}(a).  We can see that $d_{max}$ decreases with increasing of the defect depth $|\epsilon |$. We also  find similar phenomenon on the traditional defect solitons in non-PT-symmetric lattices with negative defects.

For negative defects, the linear stability analysis shows that  fundamental solitons in the semi-infinite gap are stable in their whole existence region for $W_0=0.05$. As $W_0$ increases, the stability region reduces. For $W_0=0.1$, Fig. \ref{Fig7}(c) shows that the
fundamental solitons in the semi-infinite gap are unstable in the most existence region except at the edge of the gap when $d=1$, whereas they are all unstable when $d=0.5$. For comparison, the local defect solitons in the semi-infinite gap are unstable near the edge of the gap \cite{DPT2011-OE}. All fundamental solitons in the semi-infinite gap are unstable when $W_0>0.13$.
The dipole solitons in the first gap are always unstable in the whole existence
region for any value of $W_0$.

Differing from the positive and zero defects, fundamental solitons can also exist stably in the first gap. Figures \ref{Fig8}(a) - \ref{Fig8}(c) show the power of
fundamental solitons versus propagation constant in the first gap. We can see that there exists a cutoff point of propagation constant above which the fundamental solitons can exist. This feature is similar to the case of the positive and zero defects. The cutoff point shifts toward the low propagation constant with increasing of the depth of negative defect and the value of $W_0$, and the existence region of fundamental solitons in the first gap increases too. When $W_0=0.1$, the fundamental solitons in the first gap are stable in their whole existence  region, and their stability region begins to reduce when $W_0>0.16$. When
$W_0=0.2$,  fundamental solitons is unstable in the most existence region except at the edge of the gap, as shown in Fig. \ref{Fig8}(d). Their stability region vanishes when $W_0>0.23$.

The examples for the negative defect solitons are shown in Fig. \ref{Fig9}, including the fundamental solitons in the semi-infinite gap [Figs. \ref{Fig9}(a) and \ref{Fig9}(b)], the dipole solitons in the first gap [Figs. \ref{Fig9}(c) and \ref{Fig9}(d)], and the fundamental solitons  in the first gap [Figs.  \ref{Fig9}(e) and \ref{Fig9}(f)], which correspond to the circle symbols in Fig. \ref{Fig7}(c) and Fig. \ref{Fig8}(a), respectively. Their corresponding propagations are shown in Figs. \ref{Fig10}(a)-\ref{Fig10}(f). Figures
\ref{Fig9}(a) and \ref{Fig10}(a)  show the unstable fundamental soliton nearby the edge
of the gap for $d=0.5$, whereas , the fundamental soliton nearby the edge of the gap for  $d=1$ is stable as shown in Figs. \ref{Fig9}(b) and \ref{Fig10}(b).  One can
see that dipole solitons in the first gap are unstable [Figs.  \ref{Fig10}(c) and \ref{Fig10}(d)], whereas
fundamental solitons in the first gap are stable [Figs.  \ref{Fig10}(e) and \ref{Fig10}(f)] for $W_0=0.1$.

\begin{figure}[htbp]
    \centering
    \includegraphics[width=6.7cm]{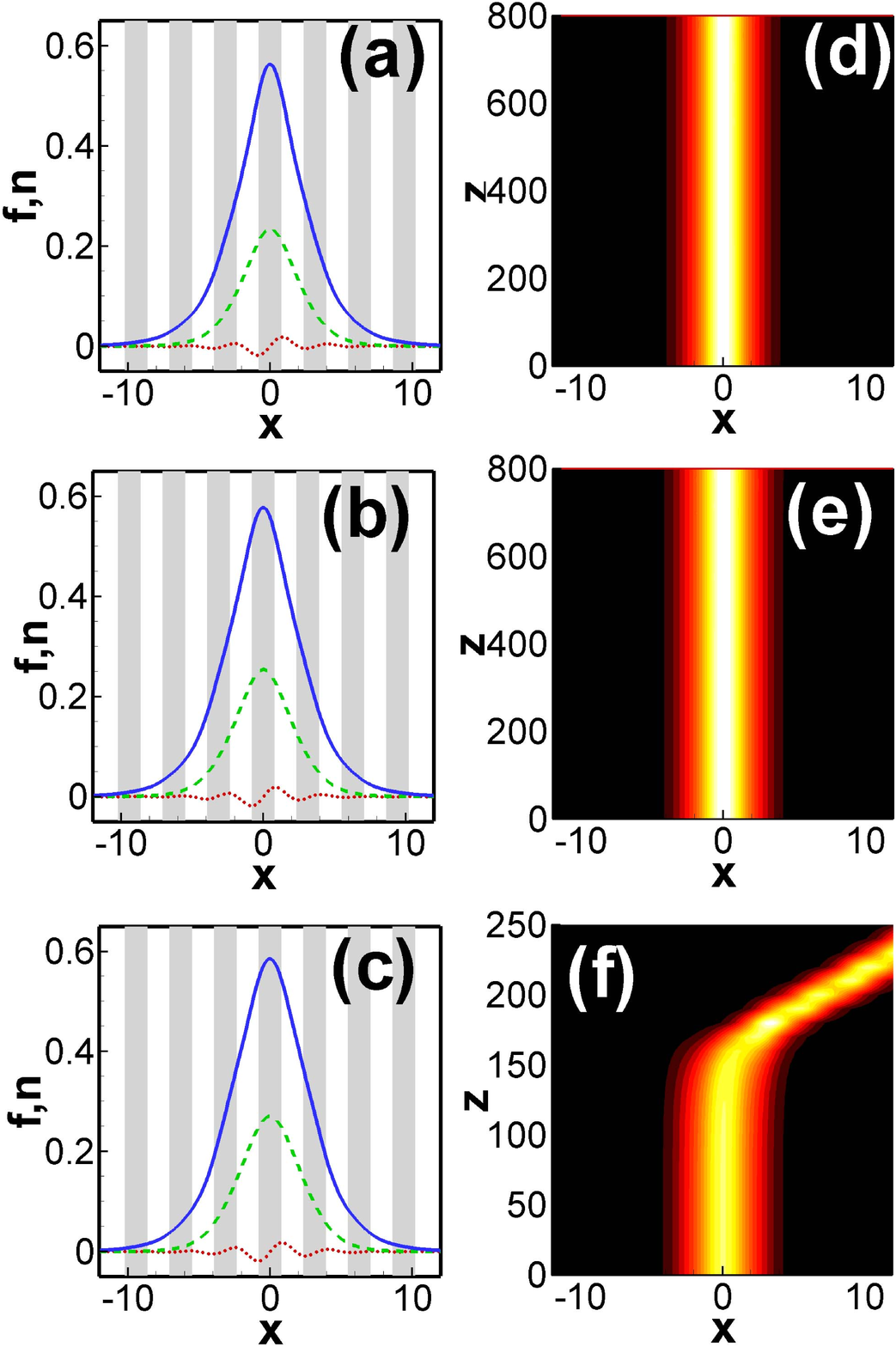}
    \caption{(color online) The complex fields (solid blue: real part; dotted red: imaginary part) and  the refractive-index changes (dashed green) of fundamental solitons  with (a) positive defect $\epsilon=0.5$,  (b) zero defect $\epsilon=0$, and (c) negative defect $\epsilon=-0.5$. (d)-(f) Evolutions of defect solitons corresponding to those in (a)-(c), respectively. For all cases $p=0.1$, $W_0=0.55$, $b=0.2$ and $d=1$.} \label{Fig11}
\end{figure}

In our study, all defect solitons are unstable for large $W_0$ when $p=4$. However, the solitons in Ref. \cite{PT2008-PRL030402} is stable when $p=1$ and $W_0=0.45$, which is close to the phase transition point ($W_0^c=0.5$). So, we finally study the solitons with small modulation strength of the PT-symmetric potentials. We find that  when the modulation depth of the PT-symmetric potentials is small, defect solitons can be stable for the positive defect and zero defect though $W_0$ is above the phase transition point. Figures \ref{Fig11}(a)-\ref{Fig11}(c) show the examples of defect solitons
above the phase transition point, i.e.  $p=0.1$ and $W_0=0.55$, with positive defects ($\epsilon=0.5$),  zero defects ($\epsilon=0$), and negative defects ($\epsilon=-0.5$), respectively.
Figures \ref{Fig11}(d)-\ref{Fig11}(f) show the corresponding propagations of those solitons. With the supporting of the nonlocal nonlinearity ($d=1$),  defect solitons with positive and zero defects are stable when $W_0$ is above the phase transition point, whereas they are unstable for negative defects.
Here the self-guiding effect by the nonlocal nonlinearity plays more important role than the defect guiding effect, in balance with the diffraction and the transverse energy flowing
induced by the imaginary part of Pt-symmetric lattice. In an other word, the nonlocal nonlinearity enhanced the real part of the PT potential, thus the nonlinearity can change the phase transition point equivalently \cite{PT2008-PRL030402}.

\section{Summary}
In conclusion, we have studied the existence and stability of defect
solitons supported by PT-symmetric nonlocal optical lattices. It is
found that the nonlocality expands the stability region of defect
solitons, especially in the first gap.
Fundamental solitons can exist stably in the semi-infinite gap for all kinds of defects
or in the first gap for negative defects, whereas dipole solitons exist in the first gap for  all kinds of defects.
Most of defect solitons are stable for small value of $W_0$, and their stability regions reduce and finally vanish as increasing $W_0$. As an exception,  dipole solitons with negative defects are always unstable for any value of $W_0$. It is interesting that there exists a maximum nonlocality degree $d_{max}$ for some negative defect solitons.
We also find defect solitons with positive and zero defects can be stable even if the PT-symmetric potentials is above the phase transition point when the modulation depth of the PT-symmetric lattices is small.
These properties of the defect solitons in PT-symmetric nonlocal lattices are obviously different from those in the local PT-symmetric lattices.  The various types of solitons
may provide alternative methods in potential applications of synthetic PT-symmetric systems.

\section*{Acknowledgments} This research was supported by the
National Natural Science Foundation of China (Grant Nos. 10804033, 11174090
and 11174091).

%%%%%%%%%%%%%%%%%%%%%%% References %%%%%%%%%%%%%%%%%%%%%%%%%


\begin{thebibliography}{99}
\bibitem{Bender-1999}
C. M. Bender, D. C. Brody, and H. F. Jones,
%``PT -symmetric quantum mechanics,''
J. Math. Phys. {\bf 40}, 2201 (1999).

\bibitem{Bender-2003}
C. M. Bender, D. C. Brody, and H. F. Jones,
%``Must a Hamiltonian be Hermitian?''
Am. J. Phys. {\bf 71}, 1095 (2003).

\bibitem{El-Ganainy2007-ol}
R. El-Ganainy, K. G. Makris, D. N. Christodoulides, and Ziad H. Musslimani
%"Theory of coupled optical PT-symmetric structures"
\ol  {\bf 32}, 2632 (2007).

\bibitem{PT2008-PRL030402}
Z. H. Musslimani, K. G. Makris, R. El-Ganainy, and D. N. Christodoulides,
% "Optical solitons in PT periodic potentials,"
\prl  {\bf 100}, 030402 (2008).

\bibitem{PT2008-PRL103904}
K. G. Makris, R. El-Ganainy, D. N. Christodoulides, and Z. H. Musslimani,
% "Beam Dynamics in PT symmetric Optical Lattices,"
\prl  {\bf 100}, 103904 (2008).


\bibitem{PT2010-PRL}
C. T. West, T. Kottos, and T. Prosen,
%``PT-Symmetric Wave Chaos,''
\prl   {\bf 104}, 054102 (2010).


\bibitem{PT2009-PRL}
O. Bendix, R. Fleischmann, T. Kottos, and B. Shapiro,
%``Exponentially fragile PT symmetry in lattices with localized eigenmodes,''
\prl  {\bf 103}, 030402 (2009).

\bibitem{New2011-pra}
A. E. Miroshnichenko, B. A. Malomed, and Y. S. Kivshar,
%"Nonlinearly PT-symmetric systems: Spontaneous symmetry breaking and transmission resonances",
\pra   {\bf 84}, 012123 (2011).

\bibitem{PT2009-prl093902}
A. Guo, G. J. Salamo, D. Duchesne, R. Morandotti, M. Volatier-Ravat,
V. Aimez, G. A. Siviloglou, and D. N. Christodoulides,
%``Observation of PT-symmetry Breaking in Complex Optical Potentials,''
\prl {\bf 103}, 093902 (2009).


\bibitem{nature-phys2010}
C. E. Ruter, K. G. Makris, R. El-Ganainy, D. N. Christodoulides, M.
Segev, and D. Kip
% "Observation of parity-time symmetry in optics"
Nature Phys. {\bf 6}, 192 (2010).


\bibitem{PT2010-Zheng}
M. C. Zheng, D. N. Christodoulides, R. Fleischmann, and T. Kottos,
%``PT optical lattices and universality in beam dynamics,''
\pra   {\bf 82},  010103 (2010).

\bibitem{New-PRA2010}H. Ramezani, T. Kottos, R. El-Ganainy, and D. N. Christodoulides,
%Unidirectional nonlinear PT -symmetric optical structures,
\pra   {\bf 82}, 043803 (2010)

\bibitem{PT2010-Sukhorukov}
A.  A. Sukhorukov, Z. Y. Xu, and Y. S. Kivshar,
%``Nonlinear suppression of time reversals in PT -symmetric optical couplers,''
\pra   {\bf 82}, 043818 (2010).


\bibitem{PT2011-Zin}
Z. Lin, H. Ramezani, T. Eichelkraut, T. Kottos, H. Cao, and D. N. Christodoulides,
%``Unidirectional invisibility induced by PT-symmetric periodic structures,''
\prl   {\bf 106}, 213901 (2011).

\bibitem{PT2011-Abdullaev}
F.Kh. Abdullaev, Y. V. Kartashov, V.V. Konotop, and D. Zezyulin,
%``Solitons inPT symmetric nonlinear lattices,''
\pra   {\bf 83}, 043805(R) (2011).

\bibitem{PT2011-Zezyulin}
D. A. Zezyulin, Y. V. Kartashov, and V.V. Konotop,
%``Stablity of solitons in PT-symmetric nonlinear potentials,''
Europhysics Letters   {\bf 96}, 64003 (2011).

\bibitem{DS2011-ol}
X. Zhu, H. Wang,  L. X. Zheng, H. G. Li, and Y. J. He,
%``Gap solitons in parity-time complex periodic optical lattices with the real part of superlattices,''
\ol  {\bf 36},  2680 (2011).


\bibitem{OL2010-Zhou}
K. Y. Zhou, Z. Y. Guo, J. C. Wang and S. T. Liu,
%``Defect modes in defective parity-time symmetric periodic complex potentials ,''
\ol   {\bf 35},  2928 (2010).


\bibitem{DPT2011-OE}
H. Wang and J. Wang,
%``Defect solitons in parity-time periodic potentials,''
Opt. Express   {\bf 19},  4030 (2011).

\bibitem{Lu2011-OE}
Z. E. Lu, and Z. M. Zhang
%``Defect solitons in parity-time symmetric superlattices,''
Opt. Express {\bf 19}, 11457 (2011).

\bibitem{DS2008-PRA}
F. Ye, Y. V. Kartashov, V. A. Vysloukh, and L. Torner,
%``Nonlinear switching of low-index defect modes in photonic lattices,''
\pra   {\bf 78}, 013847 (2008).

\bibitem{DS2009-pra}
Y.  Li, W. Pang, Y.  Chen, Z.  Yu, J. Zhou,  and H.  Zhang,
%'`Defect-mediated discrete solitons in optically induced photorefractive lattices,''
\pra   {\bf 80}, 043824 (2009).

\bibitem{DS2006-prl}
 V. A. Brazhnyi, V. V. Konotop, and V. M. Perez-Garcia,
 %``Driving defect modes of Bose-Einstein  condencates in optical lattices,''
 \prl   {\bf 96}, 060403 (2006).

\bibitem{DS2005-prl}
F. Fedele, J. Yang, and Z. Chen,
%``Defect modes in one-dimensional photonic lattices,''
\ol   {\bf 30}, 1506 (2005).

\bibitem{DS2004-PHT}
 D. K. Campbell, S. Flach, and Y. S. Kivshar,
% ``Localizing energy through nonlinearity and discreteness,''
 Phys. Today  {\bf 57},  43 (2004).



\bibitem{Segev1992-prl} M. Segev, B. Crosignani, A. Yariv, and B. Fischer,
%¡°Spatial solitons in photorefractive media,¡±
\prl    {\bf 68}(7), 923  (1992).

\bibitem{Segev1994-prl} M. Segev, G. C. Valley, B. Crosignani, P. DiPorto, and A. Yariv,
%"Steady-state spatial screening solitons in photorefractive materials with external applied  field,¡±
\prl   {\bf 73}(24), 3211  (1994).

\bibitem{Segev1998-prl} W. Krolikowski, M. Saffman, B. Luther-Davies, and C. Denz,
%¡°Anomalous interaction of spatial solitons in photorefractive media,¡±
\prl    {\bf 80}(15), 3240  (1998).


\bibitem{conti2003-prl}C. Conti, M. Peccianti, and G. Assanto,
%¡°Route to nonlocality and observation of accessible solitons,¡±
\prl  {\bf 91}, 073901 (2003).

\bibitem{Conti2004-prl}C. Conti, M. Peccianti, and G. Assanto,
%¡°Observation of optical spatial solitons in a highly nonlocal medium,¡±
\prl  {\bf 92}, 113902 (2004).


\bibitem{Peccianti2002-OL}
M. Peccianti, K. A. Brzdakiewicz, and G. Assanto,
%``Nonlocal spatial soliton interactions in nematic liquid crystals,''
\ol   {\bf 27}, 1460  (2002)

\bibitem{Rotschild2005-prl} C. Rotschild, O. Cohen, O. Manela, M. Segev, and T. Carmon
%¡°Solitons in nonlinear media with an infinite range of nonlocality:
%first observation of coherent elliptic solitons and of vortex-ring solitons,¡±
\prl   {\bf 95}, 213904 (2005).

\bibitem{Alfassi2007-prl}B. Alfassi, C. Rotschild, O. Manela, M. Segev, and D. N. Christodoulides,
%¡°Nonlocal surface-wave solitons,¡±
\prl   {\bf 98}, 213901 (2007).

\bibitem{Science1997-Snyder}
A.W. Snyder and D. J. Mitchell,
%``Accessible solitons,''
Science   {\bf 276}, 1538 (1997).

\bibitem{prl2005-Xu}
Z.  Xu, Y. V. Kartashov, and L. Torner,
%``Soliton mobility in nonlocal optical lattices,''
\prl    {\bf 95}, 113901 (2005).


\bibitem{Buccoliero2007-PRL}
D. Buccoliero, A. S. Desyatnikov, W. Krolikowski, and Y. S. Kivshar,
%``Laguerre and Hermite soliton clusters in nonlocal nonlinear media,''
\prl   {\bf 98}, 053901 (2007).


\bibitem{Conti2007-prl} N. Ghofraniha, C. Conti, G. Ruocco and S. Trillo,
%¡°Shocks in nonlocal media,¡±
\prl   {\bf 99}, 043903 (2007).

\bibitem{Conti2009-PRL}
C. Conti, A. Fratalocchi, M. Peccianti, G. Ruocco, and S. Trillo,
%``Observation of a gradient catastrophe generating solitons,''
 \prl    {\bf 102}, 083902 (2009).


\bibitem{Hu2006-APL}
W. Hu, T. Zhang, Q. Guo, X. Li, and S. Lan,
%``Nonlocality-controlled interaction of spatial solitons in nematic liquid crystals,''
\apl   {\bf 89}, 071111 (2006).



\bibitem{Nikolov2003-prl}N. I. Nikolov, D. Neshev, O. Bang, and W. Z. Krolikowski,
%¡°Quadratic solitons as nonlocal solitons,¡±
\pre   {\bf 68}, 036614 (2003).

\bibitem{Larsen2006-PRE}
P. V. Larsen, M. P. Sorensen, O. Bang, W. Z. Krolikowski, and
S. Trillo,
%``Nonlocal description of X waves in quadratic nonlinear materials,''
 \pre    {\bf 73}, 036614 (2006).

\bibitem{newnonlocal}
R. Bekenstein and M. Segev,
%"Self-accelerating optical beams in highly nonlocal nonlinear media,
Opt. Express {\bf 19}, 23706-23715(2011).


\bibitem{yang-2007}
J. Yang, and T. I. Lakoba, Stud. Appl. Math. {\bf 118,} 153-197
(2007).


\end{thebibliography}
\end{document}